%% file: paper.tex
\documentclass[conference]{IEEEtran}

\newcommand{\ignore}[1]{}
\usepackage{fancyhdr}
\usepackage[normalem]{ulem}

\newcommand{\code}[1]{\texttt{#1}}

\usepackage{listings}
\usepackage{color}
\usepackage{algpseudocode}
\usepackage{blindtext}
\usepackage{color}   % \color
\usepackage{amssymb,amsmath}
\usepackage{comment}
\usepackage{enumerate}
\usepackage{framed}
\usepackage{array}      % table tools
\usepackage{xspace}
\usepackage{multirow}
\usepackage{listings}
\usepackage{booktabs}
\usepackage{amsmath}
\usepackage{datetime}
\usepackage{subcaption}
\usepackage{graphicx}
\usepackage{microtype}
\usepackage{hyperref}
\usepackage{balance}
\usepackage{microtype}
\usepackage{dblfloatfix}

\usepackage{mathtools}

\DeclarePairedDelimiter\abs{\lvert}{\rvert}
\DeclareMathOperator{\enc}{enc}
\DeclareMathOperator{\dec}{dec}
\graphicspath{{figures/}{pics/}}
\DeclareGraphicsExtensions{.pdf,.jpeg,.png}

\title{Synthesizing Number Generators for Stochastic Computing using Mixed Integer Programming}
\author{Vincent T. Lee, Archibald Samuel Elliot, Armin Alaghi, Luis Ceze\\University of Washington}

\begin{document}
\sloppy
\maketitle
\pagestyle{plain}
\bstctlcite{IEEEexample:BSTcontrol}

%%%%%% -- PAPER CONTENT STARTS-- %%%%%%%%

\input{00-abstract}
\input{01-introduction}
\input{02-background}
\input{03-approach}
\input{04-evaluation}
\input{07-related-work}
\input{08-conclusion}

%%%%%%% -- PAPER CONTENT ENDS -- %%%%%%%%

%%%%%%%%% -- BIB STYLE AND FILE -- %%%%%%%%

\balance
\bibliographystyle{IEEEtranS}
\bibliography{references}

%%%%%%%%%%%%%%%%%%%%%%%%%%%%%%%%%%%%

\end{document}

%% file: 00-abstract.tex
\begin{abstract}

  Stochastic computing (SC) is a high density, low-power computation technique which encodes values as unary bitstreams instead of binary-encoded (BE) values.
  Practical SC implementations require deterministic or pseudo-random number sequences which are optimally correlated to generate bitstreams and achieve accurate results.
  Unfortunately, the size of the search space makes manually designing optimally correlated number sequences a difficult task.
  To automate this design burden, we propose a synthesis formulation using mixed integer programming to automatically generate optimally correlated number sequences.
  In particular, our synthesis formulation improves the accuracy of arithmetic operations such as multiplication and squaring circuits by up to $2.5\times$ and $20\times$ respectively.
  We also show how our technique can be extended to scale to larger circuits.

\end{abstract}

%% file: 01-introduction.tex
\section{Introduction}

Stochastic computing (SC) is an emerging computation technique which offers a low power, compact, and error tolerant alternative to conventional binary-encoded (BE) computation.
In SC, values are encoded as unary bitstreams (time series of 1s and 0s).
The value $p_X$ of a bitstream $X$ is interpreted as the sum over the weights of each position in the bitstream divided by the bitstream length.
For instance, the bitstream $X = 10100100$ has value $p_X = 0.375$ if we weight 1s as $+$1 and 0s as 0.
This unary encoding allows for compact implementations of arithmetic operations such as multiplication.
For instance, given two input bitstreams $X$ and $Y$, a multiplier can be implemented using a two-input AND gates since the resulting bitstreams $Z$ has value $p_Z = p_Xp_Y$.
The multiplier is only accurate if $X$ and $Y$ are uncorrelated otherwise the fundamental assumption that $p_Z = p_X \wedge p_Y = p_X p_Y$ does not hold.

To generate bitstreams, we use stochastic number generators (SNGs).
An SNG is realized by a comparator that takes a number sequence and compares it against the target value to be encoded.
Original definitions of SC assume number sequences are generated using purely random noise sources~\cite{gaines69}.
However, practical implementations of SC use deterministic or pseudo-random sources which yield more accurate results than purely random noise sources~\cite{low-discrepancy-sequences, sc-sobol}.
The selection of number sequences for SNGs is important since correlated or uncorrelated number sequences will generate correlated or uncorrected bitstreams respectively.
The correlation between bitstreams governs the functionality and accuracy of arithmetic operations in SC, and each arithmetic operation in SC has a correlation under which it is most accurate.

The key challenge is that manually engineering optimally correlated SNGs requires exploration of an exponentially large design space.
For instance, exhaustively searching through all number sequences of length of 16 would mandate evaluating $16!\approx 20$ trillion potential number sequences.
As a result, existing work relies on a handful of number sequences with desirable correlation properties and reasonable implementation costs such as low discrepancy sequences, linear feedback shift registers (LFSRs), and pulse-width modulated analog signals~\cite{Najafi17PWM}.
This leaves a large space of number sequences which may yield more accurate results than known number sequence combinations.

This paper proposes an automated method for synthesizing SNG number sequences using mixed integer programming to synthesize accurate arithmetic operations.
Our synthesis formulation generates optimally accurate number sequences for arithmetic operations such as multiplication and squaring.
More importantly, our technique eliminates the design burden of selecting properly correlated number sequences for SC circuits.
Our synthesis formulation can also be extended to synthesize RNG number sequences for larger circuits.

Our contributions are as follows:
(1) A general mixed integer program formulation for synthesizing optimal SNG number sequences.
(2) More accurate SC multiplication, and squaring circuits using synthesized number sequences.
(3) Synthesis extensions to improve scalability to larger circuits.

%% file: 02-background.tex
\section{Background}
\label{sec:background}

\noindent This section provides background on stochastic computing, impact of correlation, and the role of SNGs.

\subsection{Stochastic Computing Basics}

\noindent Stochastic computing (SC) is a computing technique proposed in the 1960s by Gaines~\cite{gaines69}.
Unlike binary-encoded (BE) computation, SC relies on unary bitstreams (time series of 1s and 0s) or stochastic numbers (SNs) to encode values.
Given an SN $X$, the encoded value of the SN $p_X$ is defined as the sum over each position in the SN length divided by the total length of the SN $N$.

SNs typically either use unipolar or bipolar encodings.
In unipolar encodings, zeros in the SN are weighted as 0 and ones are weighted as $+$1; this limits unipolar encodings to the positive range $[0, 1]$.
For instance, the SN $X = 10100000$ has value $p_X = 0.25$ since there are six 0s, two 1s, and the SN length $N = 8$.
In contrast, bipolar encodings weight zeros as $-$1 and ones as $+$1, allowing them to encode the range $[-1, +1]$.
For example, the same SN $X = 10100000$ has value $p_X = -0.5$.

\begin{figure}[t]
  \centering
  \includegraphics[width=\linewidth]{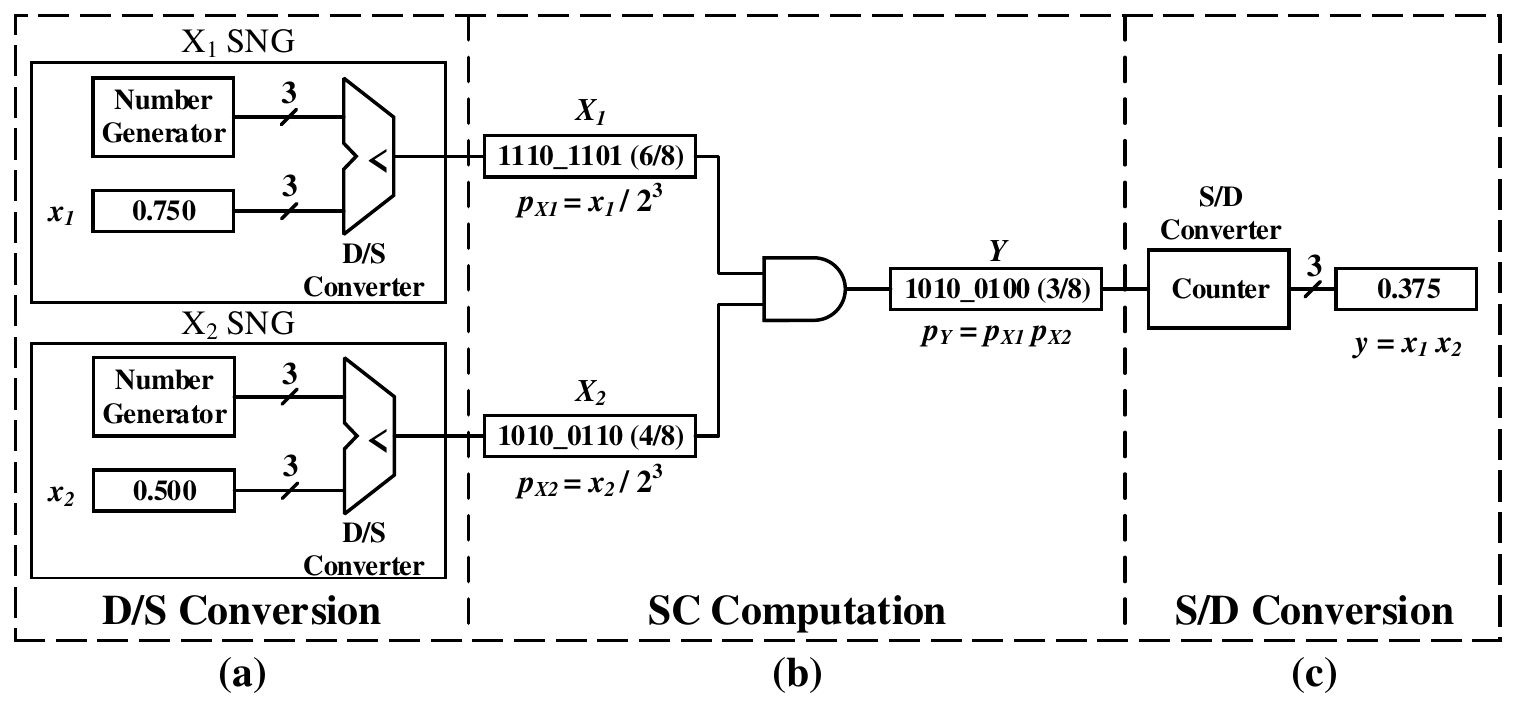}
  \caption{Example of SC multiplication: (a) D/S conversion using SNGs generates SNs, (b) SC computation, and (c) S/D conversion converts SNs back to BE values.}
  \label{fig:sc_example}
\end{figure}

A typical end-to-end computation pipeline using SC is shown in \autoref{fig:sc_example}.
In order to generate SNs, BE values are first processed by a stochastic number generator (SNG) which consists of a number sequence generator and digital-to-stochastic (D/S) converter.
The number sequence generator is responsible for producing the number sequence that drives digital-to-stochastic (D/S) converters to generate the SN.
D/S converters are commonly implemented by a comparator which takes the target BE value and the number sequence generator output to construct an SN of the target value.
It is worth noting that the comparators can be replaced by other probability shaping circuitry (e.g., chain of multiplexors) with equivalent results. 
Once SNs are generated, they can be used to perform arithmetic in the SC domain before they are converted back to the BE domain.
To convert from SC to BE, we use stochastic-to-digital (S/D) converters which are implemented by a counter.

Unipolar multiplication in SC is implemented using a two-input AND gate (\autoref{fig:sc_example}b).
Notice that the multiplication is only accurate because the inputs are uncorrelated.
Using correlated input SNs results in errors which illustrates one of the principles design challenges of SC: mitigating errors due to unfavorable correlation (discussed next).
Finally, what SC gains in lower power and higher density, it loses in terms of run time since SNs take multiple cycles to process.

The choice of number sequence when generating an SN is important as it governs the initial correlation between SNs.
Two SNs generated using the same number sequence will be positively correlated while two SNs generated from uncorrelated number sequence will be uncorrelated.
Since number sequence generators are more expensive than individual SC arithmetic operations designers often amortize their cost by exploiting application data reuse and generating many SNs from the same SNG~\cite{ichihara-iccd14}.

\subsection{Correlation and the Role of Number Sequences}

\noindent Correlation between SNs is one of the principle sources of errors in SC.
Most SC circuits have an input SN correlation under which they function optimally.
For instance, the SC multiplier introduced earlier has an affinity for uncorrelated SNs, otherwise the assumption that $p_Z = p_X \wedge p_Y = p_X p_Y$ does not hold and the computation results in errors.
Correlation between bitstreams is measured using the stochastic computing correlation (SCC)~\cite{sc-correlation}.
Given two bitstreams, an SCC of $+$1.0 indicates maximum positive correlation, an $-$1.0 indicates maximum negative correlation, and an SCC of 0.0 indicates uncorrelated.
In general, the closer the correlation of the input bitstreams are to the optimal SCC, the more accurate the computation will be.

As a result, managing and mitigating the impact of unwanted correlation between SNs is a key design challenge in SC.
There are three principle methods for controlling correlation in SC: (1) using correlation insensitive circuits, (2) using correlation manipulating circuits~\cite{lee18-date}, and (3) judiciously selecting number sequences for SNGs (this work).
The first method - using correlation insensitive circuits - relies on special variants of each arithmetic circuit which are immune to correlation levels such as the adder proposed in~\cite{lee17-date}.
Correlation insensitive circuits trade higher power and area for accuracy, and do not exist for all SC arithmetic operations.
The second correlation engineering technique is to insert correlation manipulating circuits between arithmetic operations such as isolators~\cite{ting16}, synchronizers, desynchronizers, decorrelators~\cite{lee18-date}, and regenerators.
Correlation manipulating circuits are primarily used to change correlation of existing SNs.

The third correlation engineering technique is to judiciously select number sequences for SNGs.
Recent work has shown that low discrepancy sequences such as Van der Corput (VDC), Halton~\cite{low-discrepancy-sequences}, and Sobol~\cite{sc-sobol} sequences are desirable since they have good correlation properties.
Other useful sequences include linear feedback shift registers (LFSRs), and ramp sequences~\cite{lee18-date, ramp-compare}. 
In addition, there are several unconventional SNGs such as pulse-width modulated signals~\cite{Najafi17PWM}, rotated bitstreams~\cite{jenson16}, and pre-generated bitstreams~\cite{Braendler2002}.
Beyond known techniques and sequences, there are few general methods for designing deterministic number sequences for SC circuits.
Automatically synthesizing number sequences for a target function and circuit in SC is the focus of this paper.

Finally, for combinational SC circuits, it is also possible to rotate or swap number positions within each number sequence to yield equivalently correlated bitstreams and identical computation accuracy.
For instance, two number sequences $S_X = {0, 1, 2, 3}$ and $S_Y = {0, 3, 1, 2}$ would produce \emph{iso-accurate} results as the number sequences $S_X' = {3, 0, 1, 2}$ and $S_Y' = {2, 0, 3, 1}$ (rotated versions).
We can also swap the number positions (ex. $S_X'' = {0, 2, 1, 3}$ and $S_Y'' = {0, 1, 3, 2}$) and obtain equivalently accurate results.
As long as the relative position of numbers between the number sequences are preserved, the sequences produce iso-accurate results since correlation between the number sequences is preserved.
We refer to this property as \textit{relative ordering invariance} between two number sequences which will become important later.
This property does not hold for sequential circuits since state elements are sensitive to autocorrelation which is not preserved under these transformations.

%% file: 03-approach.tex
\section{Synthesis Formulation}
\label{sec:approach}

\noindent This section outlines our mixed integer programming (MIP) problem formulation used to synthesize number sequences.

\subsection{Problem Formulation}

\noindent Linear Programming (LP) is an optimization technique that models a problem as a set of linear constraints over symbolic variables, and a linear objective function.
We use a variant known as Mixed Integer Programming (MIP) where variables are restricted to be either integer or real-valued.
The LP solver attempts to assign values to variables so that they satisfy the constraint set.
Feasible solutions are variable assignments which satisfy all constraints; optimal solutions are feasible solutions which maximize or minimize the objective function.
If no possible solution exists, the solver returns infeasible.

Our synthesis formulation defines a MIP problem that takes two input specifications: (1) a real-valued function specification, $f(p_X, p_Y)$, and (2) a hardware specification $h(X, Y)$.
The function specification $f(p_X, p_Y)$ defines the expected value of the output SN given the input SN values.
In contrast, the hardware specification, $h(X, Y)$, is a Boolean function that specifies the behavior of the underlying hardware circuit.
Given these specifications, the goal of our synthesis formulation is to produce two integer number sequences $S_X = \{x_1, ..., x_N\}$ and $S_Y = \{y_1, ..., y_N\}$ for the SNGs of $X$ and $Y$ respectively.
Each $x_i, y_i \in \mathbb{N}$ is within the range $[0, N)$ and is unique within its sequence.
We would like $S_X$ and $S_Y$ to approximate $f(p_X, p_Y)$ when used to generate the SNs $X$ and $Y$ for the hardware circuit described by $h(X, Y)$.
For instance, to synthesize the optimal number sequences for SC multiplication using a two-input AND gate, we would specify $f(p_X, p_Y) = p_Xp_Y$ and $h(X, Y) = X \wedge Y$.
Finally, we use real-valued variables because we found the ILP solver performance to be better than when using integer values.

\begin{table}[t]
  \centering
  \caption{Constraint encodings for basic logic gates.}
  \label{tab:constraints}
  \begin{tabular}{@{}cc@{}} \toprule
    Gate & Constraint Encoding \\ \midrule
    Z = AND(X, Y) & \begin{tabular}{@{}c@{}} Z $\geq$ X + Y - 1, Z $\leq$ X, Z $\leq$ Y, 0 $\leq$ Z $\leq$ 1\end{tabular} \\
    Z = OR(X, Y) & \begin{tabular}{@{}c@{}} Z $\leq$ X + Y, Z $\geq$ X, Z $\geq$ Y, 0 $\leq$ Z $\leq$ 1\end{tabular} \\
    Z = XOR(X, Y) & \begin{tabular}{@{}c@{}} Z $\leq$ X + Y, Z $\geq$ X - Y, Z $\geq$ Y - X\\ Z $\leq$ 2 - X - Y, 0 $\leq$ Z $\leq$ 1 \end{tabular} \\
    Z = NOT(X) & Z = 1 - X, 0 $\leq$ Z $\leq$ 1 \\
    \bottomrule
  \end{tabular}
\end{table}

\input{benchmark.tex}

\subsection{Solver Constraints}

\noindent We now define the MIP constraints used in our synthesis formulation to generate number sequences.
Instead of directly synthesizing the number sequence itself, we synthesize the actual SNs that correspond to each value.
To encode the number sequences $S_X$ and $S_Y$, we define two symbolic matrices of Boolean variables denoted $X_{i, j}$ and $Y_{i, j}$ where $i$ denotes the row index and $j$ denotes the SN offset.
These two symbolic matrices will encode the number sequences for the $X$ and $Y$ SNGs and are constrained such that:
\vspace{-4mm}

\begin{equation*}
\forall i \in [0, N], j \in [0, N):\; X_{i,j} \in \{0, 1\},\; Y_{i,j} \in \{0, 1\}
\end{equation*}
  
\noindent The $i$th row of each matrix effectively encodes the SN encoding for the value $i/N$.
Under this encoding, the sum of each row must equal $i$ since, under unipolar SC representations, each position that is 1 in the SN has a weight of +1.
We refer to this set of constraint as the \textit{value constraints} which are:

\vspace{-3mm}
\begin{equation*}
\forall i \in [0, N]:\; \sum_{j=0}^{N-1} X_{i, j} = i, \; \sum_{j=0}^{N-1} Y_{i, j} = i
\end{equation*}
\vspace{-3mm}

\noindent We also introduce \textit{monotonicity constraints} which require the values in each column of $S_X$ and $S_Y$ to increase monotonically.
Suppose two SNs $X_i$ and $X_{i+1}$ encoding the values $i/N$ and $(i+1)/N$ respectively and are generated from the same number sequence $S_X$.
The key insight is that if a bit at position $n$ in $X_i$ is 1, then the bit at position $n$ must also be 1 in $X_{i+1}$.
This is because if $S_X[n] < i/N$ for a given position $n$, then it must be the case that $S_X[n] < (i+1)/N$.
Therefore:

\vspace{-4mm}
\begin{equation*}
  \forall i \in [0, N), j \in [0, N]:\; X_{i, j} \leq X_{i+1, j},\; Y_{i, j} \leq Y_{i+1, j}
\end{equation*}
\vspace{-4mm}

\noindent This constraint is a consequence of choosing comparator as the probability shaping circuit within the SNG. The monotonicity constraint combined with the value constraints enforces uniqueness in that (1) no two SNs encode the same value, and (2) each encoded number within each sequence is unique.

To encode circuit functionality, we convert the hardware specification $h(X, Y)$ into its equivalent MIP formulation and set the objective to minimize absolute error.
We assume a set of constraints $H_{X, Y}$ represents the set of \textit{hardware constraints} that enforces the hardware functionality of $h(X, Y)$.
MIP formulations of Boolean logic gates such as AND, OR, NOT, and XOR are shown in \autoref{tab:constraints}; multiplexors (MUX) use compositions of basic logic gates.
State elements like D-flip-flops (DFFs) are implemented by passing the previous cycle variable in the SN.
New indicator variables are introduced as necessary to express each constraint.
The error is captured by:

\vspace{-3mm}
\begin{equation*}
  \forall n, m \in [0, N]:\; H_{n, m} = \sum_{j=0}^{N-1} h(X_{n, j}, Y_{m, j})
\end{equation*}
\vspace{-4mm}
\begin{equation*}
\forall n, m \in [0, N]: \;
C_{n, m} = H_{n, m} - \enc(f(\dec(n), \dec(m)))
\end{equation*}
\vspace{-3mm}

\noindent Where $C_{n, m}$ captures the error between the target functionality and the resulting SNs of the synthesis formulation.
An encoding function $\enc(p_Z)$ converts the function result $p_Z = f(p_X, p_Y)$ to the number of 1-bits expected in the output SN $Z$.
Similarly, an inverse function $\dec(p_Z)$ converts the number of expected 1-bits in an SN $Z$ back to a value $p_Z$.
For unipolar circuits, $\enc(p) = N \cdot p$, because $p \in [0,1]$ whereas $H_{n,m}$ is in the range $[0,N]$.
For bipolar circuits, $\enc(p) = N \cdot (p+1)/2$ because $p \in [-1,1]$.
We then minimize the cost over the absolute error as the objective function:\footnote{Mean squared error (MSE) formulations are realized using quadratic programming but are much slower. We find that average absolute error approximates MSE well.}

\vspace{-2mm}
\begin{equation*}
  \mathrm{minimize}\; \sum_{n=0}^{N} \sum_{m=0}^{N}  \abs{C_{n, m}}
\end{equation*}
\vspace{-2mm}

\noindent The absolute value function is implemented using two auxiliary variables\footnote{An auxiliary variable is a new temporary variable.} per term.
Given a cost term $C_{n,m}$, we define two auxiliary variables $t_{n,m+}$ and $t_{n,m-}$ and impose the constraints:

\vspace{-3mm}
\begin{equation*}
  \forall n, m \in [0, N]: t_{n,m+} - t_{n,m-} = C_{n,m}
\end{equation*}
\begin{equation*}
  \forall n, m \in [0, N]: \abs{C_{n,m}} = t_{n,m+} + t_{n,m-}
\end{equation*}
\begin{equation*}
  \forall n, m \in [0, N]: t_{n,m+} \geq 0, t_{n,m-} \geq 0, C_{n, m} \geq 0
\end{equation*}
\vspace{-5mm}

\noindent If $C_{n,m}$ is positive then $t_{n,m+} = C_{n,m}$ and $t_{n,m-} = 0$, otherwise $t_{n,m+} = 0$ and $t_{n,m-} = - C_{n,m}$.
The absolute value of this cost component is then expressed as $t_{n,m+} + t_{n,m-}$.
Since we minimize over the cost terms, the solver forces either $t_{n,m+}$ or $t_{n,m-}$ to zero since other assignments would be suboptimal.

The resulting number sequences $S_X$ and $S_Y$ are recovered from the variables $X_{i,j}$ and $Y_{i,j}$ by summing over each column and subtracting the sum from $N$.
Recall that SNs are generated by taking the number sequence value $s$ and checking if it is less than the target value $x$.
If $s < x$, the D/S converter emits a $1$ otherwise it emits a $0$ which means the number of zeros is proportional to the number of values where $s < x$.
More precisely:
\vspace{-2mm}
\begin{equation*}
  \forall i \in [0, N]: S_X[i] = N - \sum_{j=0}^N X_{i, j}, S_Y[i] = N - \sum_{j=0}^N Y_{i, j}
\end{equation*}
\vspace{-4mm}

\subsection{Optimization Constraints}

\noindent We now introduce two constraint optimizations to improve the run time of the synthesis formulation.

\noindent \textbf{Initial and Final Sequences:}
Recall that the sum of each row is equal to the row index; thus the sum of row 0 must also be zero and the sum of row $N$ must be $N$.
The only way to enforce the constraints $\sum_{j=0}^{N-1} X_{0, j}$ and $\sum_{j=0}^{N-1} X_{N, j}$ is $\forall j \in [0, N): X_{0,j} = 0, X_{N, j} = 1$ since variables are either 0 or 1, and there are $N$ positions in the row.
This modestly improves solver time by reducing the number of variables.

\noindent \textbf{Relative Ordering Invariance:} Recall that, for combinational circuits, numbers in two number sequences can be rotated or swapped as long as the relative pairing of numbers is preserved.
This means there are many equivalent solutions with the same objective function value.
This can be problematic for the solver since it must expend time exploring each equivalent solution and deduce that they all have the same objective function value.

Fortunately, we can exploit relative ordering invariance to eliminate equivalent solutions by initializing one of the number sequences $S_X$ to any number sequence (ex. ramp sequence \{0, 1, 2, 3, ..., N-1\}).
This reduces the number of symbolic variables to solve for by half since it is no longer necessary to solve for $\forall i, j: X_{i, j}$.
Once a solution $S_Y$ is synthesized, we can rotate the number sequences or swap the number positions to transform them into solutions with the same correlation.

%% file: benchmark.tex
{
\begin{table*}[t]
  \centering
  \caption{Number sequence synthesis benchmarks, specifications, and average absolute error (lower is better). Synthesized solutions are as accurate or more accurate than baseline solutions. $\dag$ indicates feasible but not optimal solutions.}
    \label{tab:benchmarks}
    \begin{tabular}{@{}c|c|c|c|cccc|cccc@{}} \toprule
      \multirow{2}{*}{Function} & \multirow{2}{*}{Encoding} & Function & Hardware & \multicolumn{4}{c}{Synthesized Error (Our Work)} & \multicolumn{4}{c}{Baseline Error (Prior Work)} \\
      & & $f(p_X,p_Y)$ & $Z = h(X, Y)$ & N=16 & N=32 & N=64 & N=128 & N=16 & N=32 & N=64 & N=128 \\ \toprule
      \multirow{2}{*}{Multiplier} & Unipolar & \multirow{2}{*}{$p_Xp_Y$} & Z = AND(X, Y) &
      0.016 & 0.0092 & 0.0049$\dag$ & 0.0027$\dag$ &
      0.032 & 0.020 & 0.012 & 0.0068 \\
      & Bipolar & & Z = XNOR(X, Y) &
      0.061 & 0.035 & 0.020$\dag$ & 0.010$\dag$ &
      0.067 & 0.043 & 0.023 & 0.014 \\ \midrule
      \multirow{2}{*}{Adder} & Unipolar & \multirow{2}{*}{$\frac{(p_X+p_Y)}{2}$} & \multirow{2}{*}{Z = MUX(X, Y, R)} &
      0.016 & 0.0078 & 0.0039 & 0.0020 &
      0.016 & 0.0078 & 0.0039 & 0.0020\\
      & Bipolar & & &
      0.031 & 0.016 & 0.0078 & 0.0039 &
      0.031 & 0.016 & 0.0078 & 0.0039 \\ \midrule
      Squarer & Unipolar & $p_X^2$ & \begin{tabular}{@{}c@{}}W = DFF(X)\\Z = AND(X, W)\end{tabular} &
      0.015 & 0.0076 & 0.0036 & 0.0020$\dag$ &
      0.030 & 0.030 & 0.034 & 0.040 \\ \midrule
      \begin{tabular}{@{}c@{}}Saturating\\Adder\end{tabular} & Unipolar & \code{min}$(1, p_X+p_Y)$ & Z = OR(X, Y) &
      0.0 & 0.0 & 0.0 & 0.0 &
      0.0 & 0.0 & 0.0 & 0.0 \\ %\midrule
      %% Divider & Unipolar & $\lfloor p_X/p_Y \rfloor$ & \begin{tabular}{@{}c@{}}Z = MUX(X, W, Z)\\W = DFF(Z)\end{tabular} &
      %% 0.0 & 0.0 & 0.0 & 0.0 &
      %% 0.0 & 0.0 & 0.0 & 0.0 \\ %\midrule
      %% Square Root & U & $\sqrt{p_X}$ & &
      %% ?.?? & ?.??? & ?.??? & ?.??? &
      %% 0.12 & 0.095 & 0.072 & 0.053 \\ \midrule
      %% \multirow{2}{*}{Maximum} & Unipolar & \multirow{2}{*}{\code{max}$(p_X,p_Y)$} & \multirow{2}{*}{Z = OR(X, Y)} &
      %% 0.0 & 0.0 & 0.0 & 0.0 &
      %% 0.0 & 0.0 & 0.0 & 0.0 \\
      %% & Bipolar & & &
      %% 0.0 & 0.0 & 0.0 & 0.0 &
      %% 0.0 & 0.0 & 0.0 & 0.0 \\ \midrule
      %% \multirow{2}{*}{Minimum} & Unipolar & \multirow{2}{*}{\code{min}$(p_X,p_Y)$} & \multirow{2}{*}{Z = AND(X, Y)} &
      %% 0.0 & 0.0 & 0.0 & 0.0 &
      %% 0.0 & 0.0 & 0.0 & 0.0 \\
      %% & Bipolar & & &
      %% 0.0 & 0.0 & 0.0 & 0.0 &
      %% 0.0 & 0.0 & 0.0 & 0.0 \\
      \bottomrule
    \end{tabular}
    \vspace{-2mm}
\end{table*}

}

%% file: 04-evaluation.tex
\section{Evaluation}
\label{sec:evaluation}

\noindent This section defines methodology, presents accuracy results, and evaluates the power and area of the synthesized number sequences.

\subsection{Methodology}

\noindent We evaluate synthesis problems for known arithmetic SC circuits to verify that our synthesis formulation is correct and synthesizes more accurate number sequences for existing arithmetic operations.
Our synthesis formulation is implemented on top of IBM CPLEX version 12.8.0~\cite{ibm-cplex}.
We ran our benchmarks on Microsoft Azure F72 v2 virtual machines running Ubuntu 16.04. %, which have 72 2.7 GHz Intel Xeon Platinum 8168 processors and 144GB of RAM.
To evaluate correctness, we use the synthesized number sequences to evaluate the average absolute error across all possible input value combinations.
We compare the average absolute error against those produced by using known number sequences in prior work. % for $S_X$ and $S_Y$ for each functional unit as our baseline.

For ``difficult'' synthesis instances that take intractable amounts of time, we either relax the optimality gap $g$ or bound the computation time.
The optimality gap $g$ is an ILP solver parameter that allows it to return a feasible solution within $g$ of the estimated optimal objective function value.
For instance, setting $g = 0.05$ expresses that it is acceptable to return a solution within 5\% of optimal.
We also restrict computation times.
In these cases, the solver returns a feasible solution and the estimated optimality gap between the returned solution and the optimal solution.
Both these techniques trade off optimality for speed for ``difficult'' instances.

\subsection{Results}

\noindent We now present accuracy results for our synthesized sequences.
For SC circuits like maximum, division, and minimum, our synthesis formulation generates positively correlated results and match the known optimal solutions in the literature~\cite{sc-divider, sc-correlation}.
For saturating addition, our formulation correctly identifies maximally negatively correlated number sequences which results in no accuracy errors.
For multiplication, the synthesis formulation finds number sequences which yields in more accurate results for both unipolar and bipolar encoding cases (\autoref{tab:benchmarks}).
While our formulation optimizes average absolute error, our results are still comparable or more accurate than prior work~\cite{lee17-date} in terms of mean squared error (MSE).
Finally, we find that synthesis times generally increase exponentially with search space size.

Examples of synthesized sequences for multipliers with SN length of $N = 16$ are shown in \autoref{tab:rng_example}.
Our synthesized results for multiplication achieve better overall accuracy by $2.5\times$ over previously solutions using a ramp, Van der Corput, or Halton sequences~\cite{lee17-date} (\autoref{fig:mult_accuracy}).
Our results also show we can generate more accurate bitstreams for the squaring function by up to $20\times$ (N = 128) which is better than using existing number generators.

{
  \setlength{\abovecaptionskip}{-5pt}
  \begin{figure}[t]
  \centering
  \includegraphics[width=0.9\linewidth]{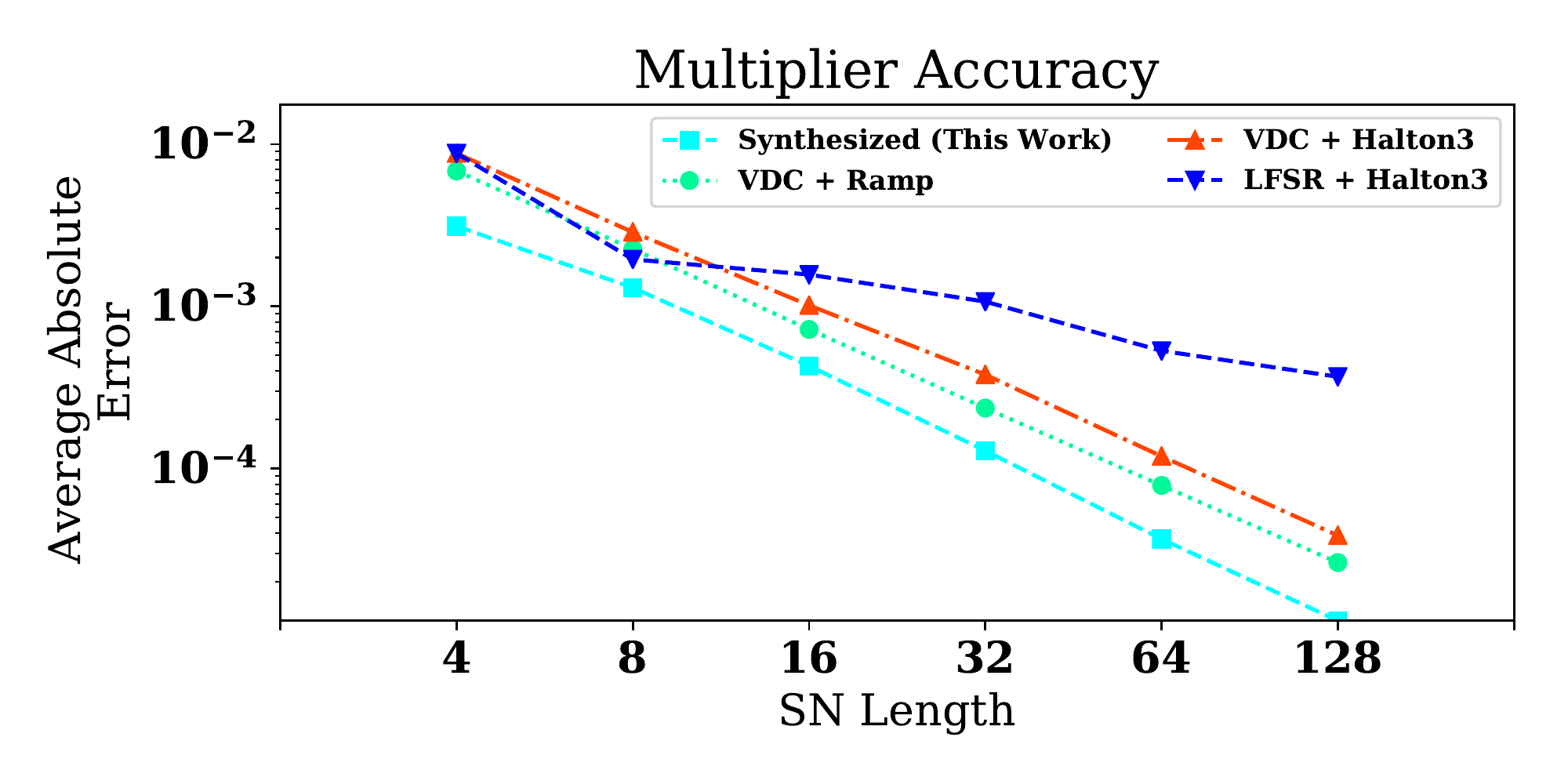}
  \caption{Multiplication accuracy. Our synthesized number sequences are optimally accurate.}
  \label{fig:mult_accuracy}
\end{figure}
}

\input{sequences.tex}

The key strength of our formulation is that provided sufficient computation resources it can \textit{automatically identify optimally correlated deterministic number sequences}.
\autoref{tab:rng_example} compares the SCC for unipolar and bipolar multiplication and shows that the \textit{average} SCC across all SNs generated by synthesized number sequences is better than using the ramp and Halton sequences proposed in~\cite{lee17-date}.

\subsection{Power, Area, and Energy}

\begin{figure}[t]
  \centering
  \includegraphics[width=0.8\linewidth]{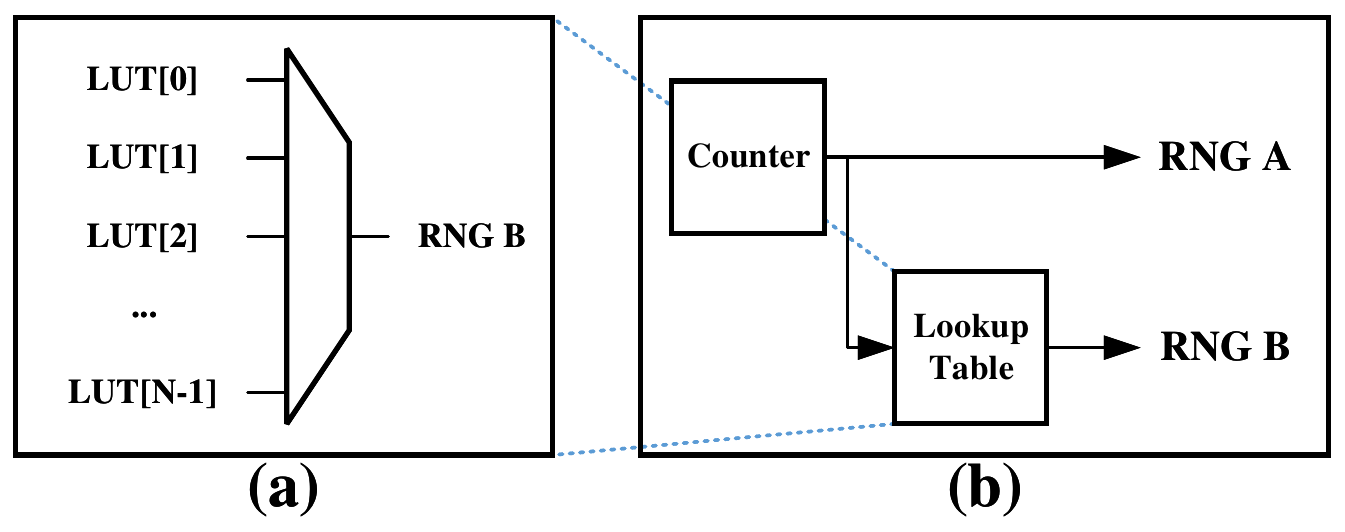}
  \caption{(a) Lookup table number generator for synthesized number sequences. (b) A counter serves as both a number generator and lookup table driver.}
  \label{fig:synth_rng}
\end{figure}

\noindent We evaluate the power, area, and energy cost of our synthesized number sequence generators by using Synopsys Design Compiler, IC Compiler, and PrimeTime Power using a 65 nm TSMC library.
We compare VDC, Halton3, and LFSR sequences with a lookup table architecture for synthesized number sequences since synthesized sequences have no obvious efficient hardware implementation.
For a two-input function, we only need one lookup table to generate $S_Y$ since we can initialize $S_X$ to a ramp function and use it to also drive the lookup table.
The architecture for this pair of number sequence generators is shown in \autoref{fig:synth_rng}.

To compare scalability, we evaluate number sequence generators for $N = 4, 8, 16, 32, 64, 128, 256$.
\autoref{fig:rng_power_area} shows the power and area comparison of several known number generators.
Compared to existing number sequence generators, individual synthesized number sequence generators consume more power and area for N=128 length SNs by up to 4.7$\times$ and 2.5$\times$ respectively; for shorter bitstream lengths, this gap quickly closes.
While the relative gaps may appear large, in the context of an end-to-end accelerator, this power and area differential has limited impact.

{
  \setlength{\belowcaptionskip}{-15pt}
  \setlength{\abovecaptionskip}{-5pt}
\begin{figure}[t]
  \centering
  \includegraphics[width=0.9\linewidth]{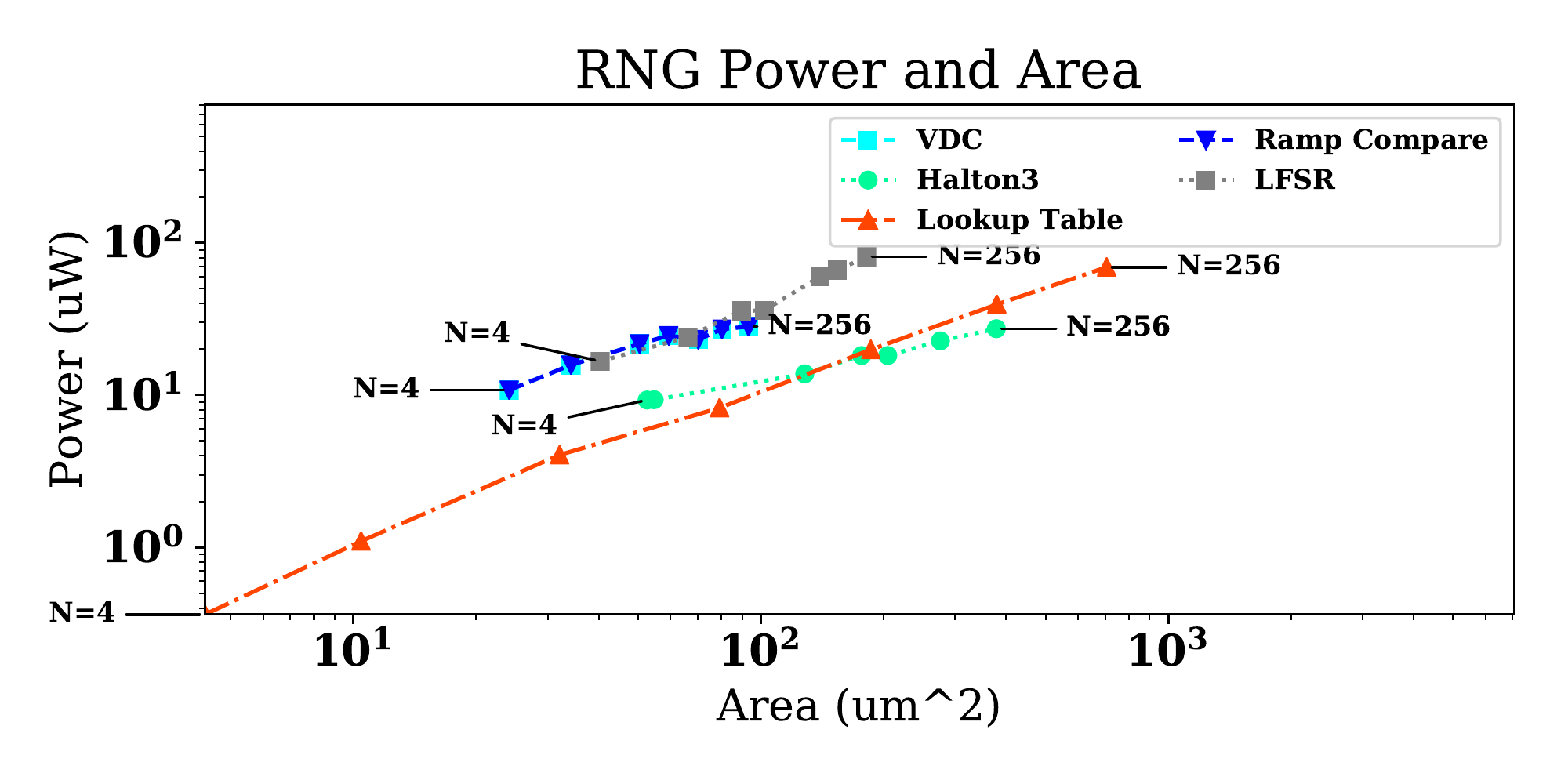}
  \caption{Power and area of number generators for N = 4, 8, 16, 32, 64, 128, 256.}
  \label{fig:rng_power_area}
\end{figure}
}

To measure the overall power and area overhead of our number generators, we evaluate a convolution and matrix-vector multiply kernel.
For convolution, we assume a 8$\times$8 input tile and 5$\times$5 kernel window.
For matrix-vector multiplication, we assume a 32$\times$32 matrix and 32 dimensional weight vector.
We measure power and area using post-placement and route results using random data inputs to make the results data agnostic.
We find that our number sequence generators consume less than 1.6\% and 1.3\% of total power and area of the end-to-end designs respectively.
Compared to designs using a ramp and VDC number generators, a design using our synthesized number generators increases the overall area by a mere 4.5\%.
This is to be expected since SC accelerators are dominated by compute, S/D, and D/S conversion overheads; a marginal increase in SNG size has minimal impact on overall power and area while improving accuracy.
Depending on the application, these energy and accuracy trade offs can be justified.

Finally, we find that synthesis times scale poorly with SN length since each additional degree of freedom doubles the search space size.
In practice, solver synthesis times are faster than worst case exponential times because solvers prune away large portions of the space.
We generally find that the CPLEX solver is limited to up SN lengths of up to N = 256.
However, prior work shows that SC is only more energy efficient at low operating precisions (N $\leq$ 256)~\cite{lee17-date} so the scalability limits are not fatal to our synthesis formulation's utility.

\subsection{Multiple Input Circuits}

\noindent To scale to additional inputs or larger circuits, we decompose the synthesis problem into smaller subproblems.
The key insight is that many $N$-input circuit can be decomposed into $N-1$ smaller two-input circuits, each which have their own function specification $f_n(p_X, p_Y)$ and hardware specification $h_n(X, Y)$.

\begin{figure}[t]
  \centering
  \includegraphics[width=\linewidth]{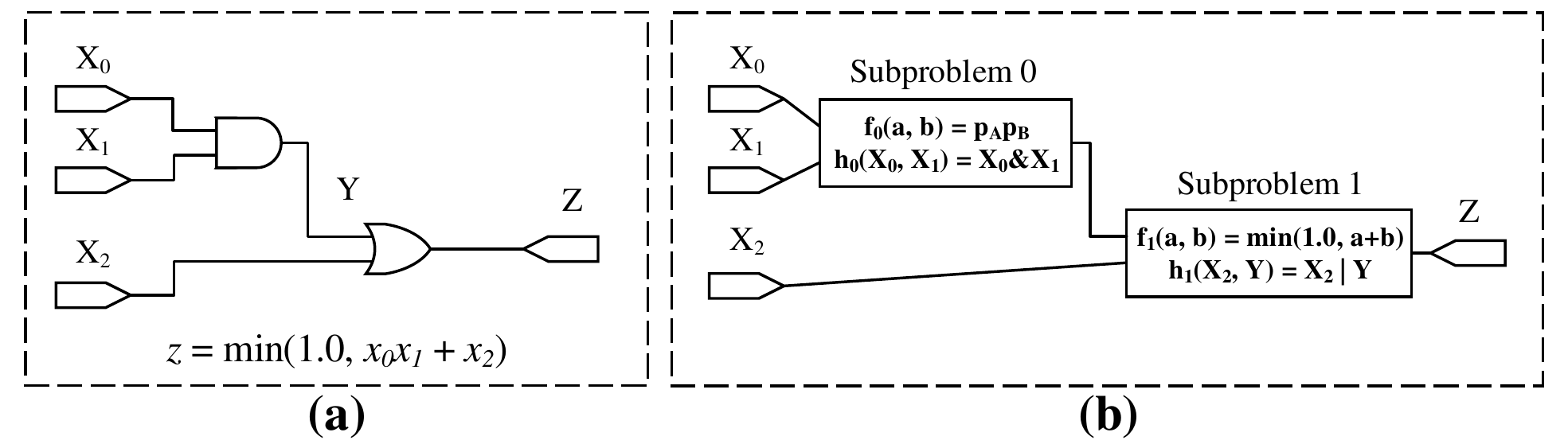}
  \caption{Circuit decomposition into subproblems. (a) Fused multiply-add with three inputs decomposed into (b) two subproblems.}
  \label{fig:decompositions}
\end{figure}

\autoref{fig:decompositions}a shows an example of how a fused-multiply add is decomposed into two subproblems (\autoref{fig:decompositions}b): one for multiplication (subproblem 0) and one for saturating addition (subproblem 1).
Notice that each subproblem encapsulates its own functionality and hardware specification.
We first synthesize the number sequences $X_0$ and $X_1$ for subproblem 0 since it occurs first in the circuit's topological order.
Given $X_0$ and $X_1$, we exhaustively generate all possible output SNs from the multiplier and record them in a $(N+1) * (N+1) \times N$ dimensional matrix $Y$.
Each row in matrix $Y$ corresponds to a possible SN output from subproblem 0.

To synthesize $X_2$ we construct a second synthesis problem.
Unlike subproblem 0, we use the output SNs in $Y$ from subproblem 0 as one of the inputs instead of a symbolic matrix corresponding to a number sequence.
We still assign a matrix of symbolic variables for $X_2$ and synthesize it.
One drawback is that the number of resulting SNs generated by subproblem 0 increases quadratically with SN length.
To mitigate this, we deduplicate the rows in $Y$; the key insight is that the first subproblem may generate redundant SNs (identical SNs).
The degree of redundancy depends on the encoded computation and hardware specification.

Decompositions present their own trade offs.
Using decompositions trades global optimality guarantees for scalability; solutions for each individual gate are still locally optimal.
By partially calculating the resulting values after subproblems 0, we eliminate the need to solve for all input number sequences at the same time which improves scalability by reducing the search space size.
Unfortunately, the synthesized number sequence results are only optimal for each subproblem and does not guarantee that the synthesized number sequences are optimal for the original circuit.

%% file: sequences.tex
\begin{table*}[t]
  \centering
  \caption{Synthesized number sequences compared to existing solutions (N=16). Our multiplier sequences on average are more optimally uncorrelated.}
  \label{tab:rng_example}
  \begin{tabular}{@{}ccccc@{}} \toprule
    Functionality & Synthesized Sequences (Our Work) & Baseline Sequences (Prior Work) &
    \begin{tabular}{@{}c@{}}Synthesized\\SCC\end{tabular} &
      \begin{tabular}{@{}c@{}}Baseline\\SCC\end{tabular} \\ \midrule %% &
        \multirow{2}{*}{\begin{tabular}{@{}c@{}}Unipolar\\Multiply\end{tabular}}
        & [0, 1, 2, 3, 4, 5, 6, 7, 8, 9, 10, 11, 12, 13, 14, 15] & [0, 1, 2, 3, 4, 5, 6, 7, 8, 9, 10, 11, 12, 13, 14, 15] &
        \multirow{2}{*}{$0.0$} & \multirow{2}{*}{$0.45$}\\
        & [6, 13, 1, 10, 8, 3, 15, 4, 11, 0, 12, 7, 5, 14, 2, 9] & [8, 4, 12, 2, 10, 6, 14, 1, 9, 5, 13, 3, 11, 7, 15, 0] & & \\
        \midrule
        \multirow{2}{*}{\begin{tabular}{@{}c@{}}Bipolar\\Multiply\end{tabular}}
        & [0, 1, 2, 3, 4, 5, 6, 7, 8, 9, 10, 11, 12, 13, 14, 15] & [0, 1, 2, 3, 4, 5, 6, 7, 8, 9, 10, 11, 12, 13, 14, 15] &
        \multirow{2}{*}{$0.0$} & \multirow{2}{*}{$0.23$}\\
        & [6, 13, 1, 10, 8, 3, 15, 4, 11, 0, 12, 7, 5, 14, 2, 9] & [0, 1, 3, 7, 15, 14, 13, 10, 5, 11, 6, 12, 9, 2, 4, 8] & & \\
        \midrule

        %% \multirow{2}{*}{\begin{tabular}{@{}c@{}}Inv. Bipolar\\Multiply\end{tabular}} & [0, 1, 2, 3, 4 5, 6, 7, 8, 9, 10, 11, 12, 13, 14, 15] & [0, 1, 2, 3, 4, 5, 6, 7, 8, 9, 10, 11, 12, 13, 14, 15] &
        %% \multirow{2}{*}{$6.9\times10^{-3}$} & \multirow{2}{*}{$8.00\times10^{-3}$}\\
        %% & [9, 2, 14, 5, 7, 12, 0, 11, 4, 15, 3, 8, 10, 1, 13, 6] & [5, 11, 2, 7, 12, 4, 9, 14, 1, 6, 11, 2, 8, 13, 4, 9] & & \\
        %% \midrule
        Square & [2, 0, 8, 12, 11, 7, 6, 1, 4, 13, 14, 5, 9, 10, 3, 15] & [0, 5, 11, 2, 7, 12, 4, 9, 14, 1, 6, 11, 2, 8, 13, 4] & -- & -- \\
        %% \midrule
        %% Square Root & [10, 1, 5, 7, 14, 4, 9, 13, 15, 0, 6, 12, 11, 8, 3, 2] & [8, 4, 12, 2, 10, 6, 14, 1, 9, 5, 13, 3, 11, 7, 15, 0] & & $1.67\times10^{-2}$ \\
        \bottomrule
  \end{tabular}
  \vspace{-3mm}
\end{table*}

%% file: 07-related-work.tex
\section{Related Work}
\label{sec:related-work}

\noindent Designing number generators for SC has typically been a manual design task that relies on designer insight.
Ichihara et al.~\cite{ichihara-iccd14} propose sharing rotated versions of LFSRs to amortize implementation cost over two SNGs.
Neugebauer et al.~\cite{neugebauer17-dsd} propose a new number sequence generator SBoNG which improves autocorrelation and cross correlation of generated SNs.
Zhakatayev et al.~\cite{even-distribution-coding} improve SNG implementation cost by using even distribution coding.
Kim et al.~\cite{kim16-aspdac} proposes an SNG that uses an auxiliary RNG to shuffle bits of an existing SN to generate a new SN.
However, most of these prior works concentrate on improving implementation cost and/or randomness of number generators, not exploring the remaining space of number generators.
To our knowledge, this work is the first to automatically synthesize optimally correlated, deterministic number sequences for SNGs.

%% file: 08-conclusion.tex
\section{Conclusions}
\label{sec:conclusions}

\noindent We presented a mixed integer program synthesis formulation for automatically generating optimally correlated number sequences to improve the accuracy of stochastic circuits.
Our formulation generalizes to any circuit and removes the design burden of identifying optimally correlation number sequences from the design process.
In particular, we show that it yields more accurate multiplication and squaring circuits, and show how it can be extended to larger multiple input circuits.